\documentclass[1p,preprint]{elsarticle}

\usepackage{hyperref}
\usepackage{siunitx}
\DeclareSIUnit\k{k}
\DeclareSIUnit\M{M}
\usepackage{xspace}
\usepackage{booktabs}
\usepackage{graphicx}
\usepackage{amsmath}
\usepackage[edges]{forest}
\usepackage{tikz}
\usepackage{tikz-3dplot} 
\usepackage{pgfplots}
\pgfplotsset{compat=1.18} 
\usepackage{pgfmath}

\usetikzlibrary{
    fit,
    math,
    calc,
    arrows,
    shapes,
    shadows,
    fadings,
    external,
    plotmarks,
    arrows.meta,
    intersections,
    pgfplots.statistics,
    pgfplots.fillbetween,
}
\usepgfplotslibrary{groupplots}
\pgfdeclarelayer{back}
\pgfsetlayers{back,main}

\newcommand{\sphenix}{\texttt{sPHENIX}\xspace}
\newcommand{\tpcpp}{\texttt{TPCpp-10M}\xspace}
\newcommand{\ppcol}{$p{+}p$\xspace}
\newcommand{\gev}[1]{\SI{#1}{\giga\electronvolt}}
\newcommand{\mev}[1]{\SI{#1}{\mega\electronvolt}}

\newcommand{\geant}{\textsc{Geant4}\xspace}
\newcommand{\pythia}{\textsc{Pythia}\xspace}
\definecolor{pion}{HTML}{66C2A5}
\definecolor{kaon}{HTML}{FC8D62}
\definecolor{proton}{HTML}{8DA0CB}
\definecolor{electron}{HTML}{E78AC3}
\definecolor{others}{HTML}{A6D854}
\definecolor{signal}{HTML}{00BFFF}
\definecolor{noise}{HTML}{FF7F50}
\newif\ifcompilefigures
\compilefiguresfalse   % use pre-generated figures
\journal{Data in Brief}

\makeatletter
\def\ps@pprintTitle{%
  \let\@oddhead\@empty
  \let\@evenhead\@empty
  \def\@oddfoot{\footnotesize\itshape
    Preprint submitted to \ifx\@journal\@empty Elsevier\else\@journal\fi\hfill\today}%
  \let\@evenfoot\@oddfoot
}
\makeatother

\begin{document}

\clearpage

\begin{frontmatter}
\title{\tpcpp: Simulated proton-proton collisions in a Time Projection Chamber for AI Foundation Models}

\author[columbia]{Shuhang Li\corref{cor1}}
\author[bnlai]{Yi Huang}
\author[bnlai]{David Park}
\author[bnlai]{Xihaier Luo}
\author[bnlphy]{Haiwang Yu}
\author[bnlphy]{Yeonju Go}
\author[bnlphy]{Christopher Pinkenburg}
\author[bnlai]{Yuewei Lin}
\author[bnlai]{Shinjae Yoo}
\author[bnlphy]{Joseph Osborn}
\author[mit]{Christof Roland}
\author[bnlphy]{Jin Huang}
\author[bnlai]{Yihui Ren}

\address[columbia]{Department of Physics, Columbia University, New York, NY, USA}
\address[bnlai]{AI Department, Brookhaven National Laboratory, Upton, NY, USA}
\address[bnlphy]{Nuclear and Particle Physics Department, Brookhaven National Laboratory, Upton, NY, USA}
\address[mit]{Department of Physics, Massachusetts Institute of Technology, Cambridge, MA, USA}

\cortext[cor1]{Corresponding author. Email: sli7@bnl.gov}
%\fntext[equal]{These authors contributed equally.} 


\begin{abstract}

Scientific foundation models hold great promise for advancing nuclear and particle physics by improving analysis precision and accelerating discovery. Yet, progress in this field is often limited by the lack of openly available large-scale datasets, as well as standardized evaluation tasks and metrics. Furthermore, the specialized knowledge and software typically required to process particle physics data pose significant barriers to interdisciplinary collaboration with the broader machine learning community.
This work introduces a large, openly accessible dataset of 10 million simulated proton–proton collisions, designed to support self-supervised training of foundation models. To facilitate ease of use, the dataset is provided in a common \texttt{NumPy} format. In addition, it includes 70,000 labeled examples spanning three well-defined downstream tasks -- track finding, particle identification, and noise tagging -- to enable systematic evaluation of the foundation model's adaptability.
The simulated data are generated using the \pythia Monte Carlo event generator at a center-of-mass energy of $\sqrt{s}=\gev{200}$ and processed with \geant to include realistic detector conditions and signal emulation in the \sphenix Time Projection Chamber at the Relativistic Heavy Ion Collider, located at Brookhaven National Laboratory.
This dataset resource establishes a common ground for interdisciplinary research, enabling machine learning scientists and physicists alike to explore scaling behaviors, assess transferability, and accelerate progress toward foundation models in nuclear and high-energy physics.
The complete simulation and reconstruction chain is reproducible with the \sphenix software stack. All data and code locations are provided under Data Accessibility.

% This article presents a large-scale dataset of 10 million simulated proton-proton collisions at center of mass energy $\sqrt{s}=\gev{200}$ in the \sphenix Time Projection Chamber (TPC) at Relativistic Heavy Ion Collider (RHIC), generated using \pythia Monte Carlo (MC) event generator and simulated using \geant with realistic detector conditions. The dataset comprises two parts: (i) an unlabeled set of \SI{10}{\M} events containing reconstructed TPC spacepoint positions (in \si{\cm}) suitable for self-supervised pretraining of foundation models, and (ii) a labeled set of $\SI{70}{\k}$ events containing spacepoint positions together with per-spacepoint truth track identifier, particle species class, and spacepoint noise labels. Event-level content also includes MC truth particles reaching the detector. The dataset supports machine learning for track finding, noise tagging, and particle identification, with recommended evaluation metrics provided. The complete simulation and reconstruction chain is reproducible with the \sphenix software stack; all data and code locations are provided under Data Accessibility.
\end{abstract}
\end{frontmatter}

\clearpage

% --------------------------
% SPECIFICATIONS TABLE
% --------------------------
\section*{Specifications Table}
\begin{table}[ht]
    \resizebox{\textwidth}{!}{
        \renewcommand{\arraystretch}{1.2}
        \begin{tabular}{p{.35\textwidth}p{.65\textwidth}}
            \toprule
            \textbf{Subject} & Physics \\
            \textbf{Specific subject area} & Charged-particle tracking simulation in a time projection chamber (TPC) at Relativistic Heavy Ion Collider (RHIC) \\
            \textbf{Type of data \& formats} & \texttt{NumPy} compressed archive \\
            \textbf{Data collection} & proton+proton (\ppcol) collision events generated with \pythia~8.307 (Detroit tune)~\cite{m:pythia8,PYTHIA_tune} at center-of-mass energy $\sqrt{s}=\gev{200}$; simulated with \geant\cite{m:g4} (FTFP\_BERT\_HP) via the as-built \sphenix detector and $\SI{1.4}{\tesla}$ magnetic field; TPC signals digitized and reconstructed into spacepoints using \sphenix software. Labeled set includes spacepoint-to-truth associations; \\
            \textbf{Data source location} & Brookhaven National Laboratory, Upton, NY, USA (simulated under \sphenix software environment). \\
            \textbf{Data accessibility} & Repository name: \textbf{Zenodo} \\
            & Data identification number (DOI): \textbf{10.5281/zenodo.16970029} \\
            & Direct URL to data: \textbf{\url{https://doi.org/10.5281/zenodo.16970029}} \\
            & Instructions for access: Public, anonymous download; dataset provided as compressed archives; example loaders and documentation included in the repository. \\

            \textbf{Related research article} & None \\
            \bottomrule
        \end{tabular}
    }
\end{table}

{
\begin{figure}[ht]
	\centering
    \tikzsetnextfilename{fig_rhic_sphenix}
    % \tikzexternaldisable
    \ifcompilefigures
        \resizebox{\textwidth}{!}{\input{figures/rhic_sphenix}}
    \else
        \resizebox{\textwidth}{!}{\includegraphics{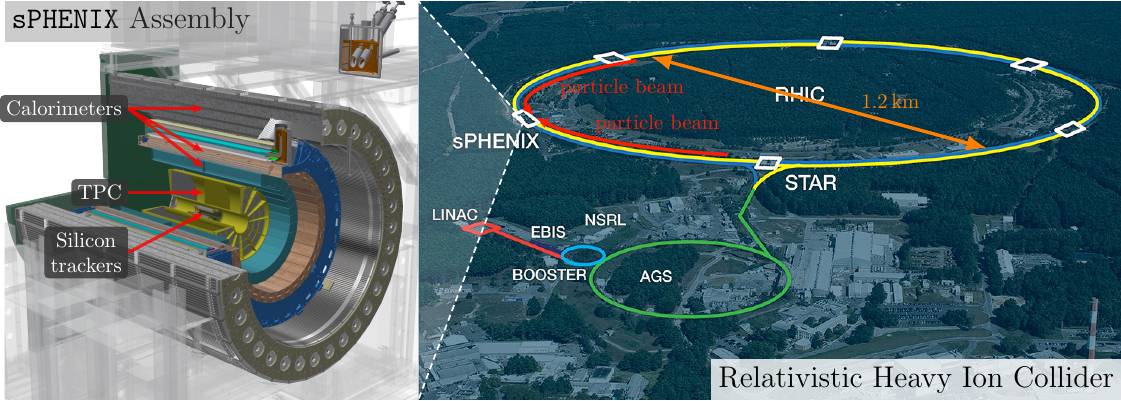}}
    \fi
    \caption{RHIC and \sphenix experiment (context figure).}
    \label{fig:rhic}
\end{figure}
}

% --------------------------
% VALUE OF THE DATA
% --------------------------
\section*{Value of the Data}
\begin{itemize}
    \item \textbf{Supports foundation model training.} The dataset provides $\num{10000000}$ unlabeled events suitable for large-scale self-supervised or contrastive pretraining, complemented by $\num{70000}$ labeled events for fine-tuning. This enables scalable applications of machine learning to charged-particle tracking.

    \item \textbf{Fills a gap in open high-energy physics datasets.} Most publicly available tracking datasets are based on silicon detectors (e.g., strip or pixel trackers) in high-energy particle collisions. In contrast, this dataset is built on a large, gas-filled TPC, the central tracker in \sphenix and many other nuclear physics experiments. By providing realistic TPC spacepoints with truth associations, it enables research on tracking challenges unique to gaseous detectors, such as long drift distances, diffusion, and complex pattern recognition across several tens of layers, compared to the $\sim$10 precise two-dimensional layers typically found in silicon trackers.

    \item \textbf{Benchmarking resource.} The labeled subset includes ground-truth information for three standard downstream tasks -- track finding, particle identification, and noise tagging -- along with recommended metrics, establishing a common benchmark for fair and reproducible evaluation.

\end{itemize}

% --------------------------
% BACKGROUND
% --------------------------
\section*{Background}

In nuclear and particle physics experiments, tracking detectors are used to measure and reconstruct the trajectories of charged particles. These measurements are crucial for determining momenta, identifying particle species, and studying underlying physics processes.
The TPC provides three-dimensional (3D) spacepoints that reconstruct the trajectories of particles traversing the detector. Figure~\ref{fig:rhic} shows the \sphenix TPC at RHIC is a cylindrical gas-filled detector with an instrumented region featuring an inner radius of $\SI{\sim 32}{\cm}$, outer radius of $\SI{\sim 78}{\cm}$, and length of $\SI{\sim 211}{\cm}$ in a $\SI{1.4}{\tesla}$ magnetic field. The TPC records the ionization signals produced as charged particles pass through the gas volume. 
%Charged-particle tracking in a solenoidal TPC provides $3$-dimensional (3D) spacepoint measurements for reconstructing particle trajectories\yi{It seems that we are using ``trajectory'' and ``track'' interchangeably. I wonder whether we should use just one.}, identifying particles, and suppressing noise. In the \sphenix experiment at RHIC, the TPC, a cylindrical gas-filled detector (inner radius $\SI{\sim 32}{\cm}$, outer radius $\SI{\sim 78}{\cm}$, length $\SI{\sim 211}{\cm}$ for the instrumented region) in a $\SI{1.4}{\tesla}$ magnetic field, records ionization signals produced by charged particles traversing the gas volume. 
Figure~\ref{fig:diagram_spacepoints} illustrates how the ionized electrons drift under a uniform electric field produced by the central membrane toward the readout planes, where their arrival times encode the longitudinal ($z$) coordinate and the channel locations in the readout planes provide the transverse ($r$–$\phi$) coordinates. In this way, each charged particle leaves a series of high-resolution spacepoints ($\SI{\sim 150}{\um}$ in $r$–$\phi$;  $\SI{\sim 750}{\um}$ in $z$) along its trajectory~\cite{sphenix_tpc}.
As the central component of tracking in \sphenix, the TPC provides precise momentum measurements from particle trajectories and particle identification.

This dataset was compiled to provide a public, high-fidelity resource of simulated minimum-bias \ppcol collisions at $\sqrt{s}=\gev{200}$, motivated by the need for openly accessible data to develop and benchmark machine learning and reconstruction algorithms for TPC tracking. Such collisions, standard at RHIC, offer a controlled environment for testing methods before tackling complex heavy-ion collision data. This dataset ensures portability and reproducibility with clearly defined units and truth associations, fostering method development across physics and data science communities without requiring proprietary software access. The large unlabeled set (\SI{10}{\M} events) is particularly well suited for training foundation models via self-supervised learning, addressing the need for scalable, high-fidelity datasets in charged-particle tracking. 

{
\begin{figure}[ht]
    \centering
    \tikzsetnextfilename{fig_diagram_spacepoints}
    % \tikzexternaldisable
    \ifcompilefigures
        \resizebox{\textwidth}{!}{
            \begin{tikzpicture}    
                \def\scale{0.0428} % I generated a diagram that is way too big ^_^
                \def\fontsz{8}
                \node[inner sep=0, draw=none] (diagram) at (0, 0) {\input{figures/tpc_diagram}};
                \node[inner sep=0, draw=none, anchor=south west] (spacepoints) at ([xshift=.06\textwidth]diagram.south east) {\includegraphics[width=.47\textwidth]{figures/trimmed_7392_data-with-coord.png}};
                \node[inner sep=0, anchor=south west] (diagram_title) at (diagram.north west) {(a) TPC schematic};
                \node[inner sep=0, anchor=south west] (spacepoints_title) at (diagram_title.south -| spacepoints.west) {(b) A collision event};
            \end{tikzpicture}
        }
    \else
        \resizebox{\textwidth}{!}{\includegraphics{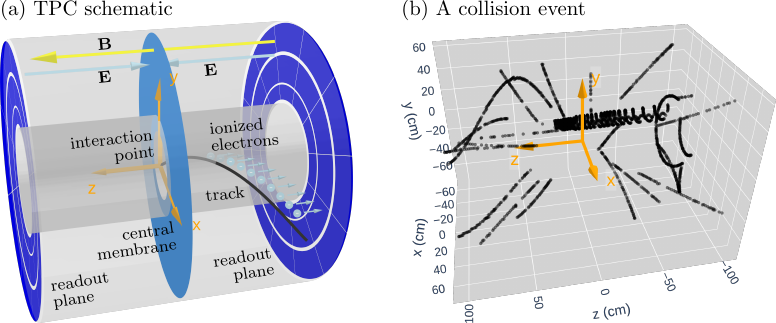}}
    \fi 
    \caption{TPC schematic and collision event.  Panel (a): The TPC schematic, where proton beams travel along the $z$ direction. The electric field $E$ drives ionization electrons to the readout planes, while the magnetic field $B$ bends charged particle trajectories. Panel (b): An event display of reconstructed spacepoints in the TPC.}
    \label{fig:diagram_spacepoints}
\end{figure}
}

% --------------------------
% DATA DESCRIPTION
% --------------------------

\section*{Data Description}

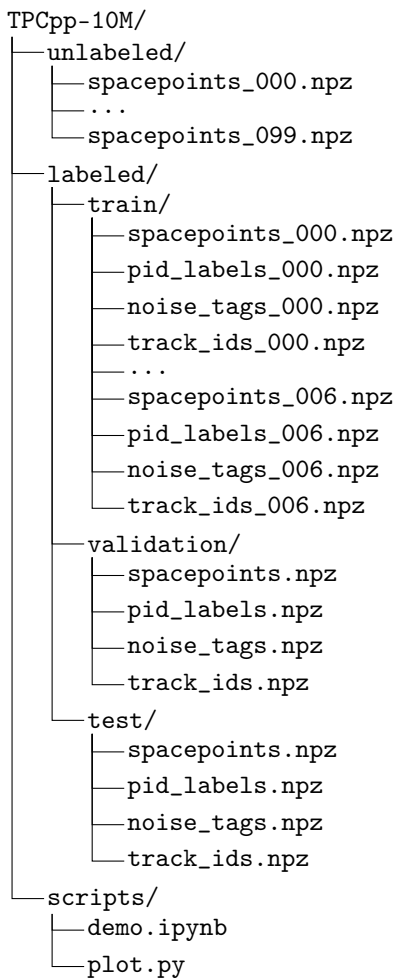
\begin{figure}[h]
    \centering
    \begin{forest}
        for tree={
            inner sep=1,
            font=\ttfamily,
            grow'=0,
            child anchor=west,
            parent anchor=south,
            anchor=west,
            calign=first,
            edge path={\noexpand\path [draw, \forestoption{edge}]
                (!u.south west) ++(3pt,0) |- (.child anchor) \forestoption{edge label};
            },
            before typesetting nodes={
                if n=1
                    {insert before={[,phantom]}}
                    {}
            },
            fit=band,
            before computing xy={l=15pt},
            s sep=3pt,  % vertical spacing between siblings
        }
        [TPCpp-10M/
            [unlabeled/
                [spacepoints\_000.npz]
                [\dots]
                [spacepoints\_099.npz]
            ]
            [labeled/
                [train/
                    [spacepoints\_000.npz]
                    [pid\_labels\_000.npz]
                    [noise\_tags\_000.npz]
                    [track\_ids\_000.npz]
                    [\dots]
                    [spacepoints\_006.npz]
                    [pid\_labels\_006.npz]
                    [noise\_tags\_006.npz]
                    [track\_ids\_006.npz]
                ]
                [validation/
                    [spacepoints.npz]
                    [pid\_labels.npz]
                    [noise\_tags.npz]
                    [track\_ids.npz]
                ]
                [test/
                    [spacepoints.npz]
                    [pid\_labels.npz]
                    [noise\_tags.npz]
                    [track\_ids.npz]
                ]
            ]
            [scripts/
                [demo.ipynb]
                [plot.py]
            ]
        ]
    \end{forest}
    \caption{Folder structure of the \tpcpp dataset.}
    \label{fig:folder-structure}
\end{figure}

The dataset repository contains simulated \ppcol collision data at 
$\sqrt{s}=\gev{200}$ for the \sphenix TPC, 
organized into three top-level folders. All data files are stored in \texttt{.npz} format, 
which can be read with Python libraries, such as \texttt{NumPy}.
The TPC is a cylindrical detector with its $z$-axis along the beam direction, centered at 
the detector origin. Spacepoint coordinates $(x, y, z)$ are reported in Cartesian coordinates (cm). 
\autoref{fig:folder-structure} provides a summary of the folder structure.

\begin{itemize}
    \item \texttt{unlabeled/}: Contains \SI{10}{\M} minimum-bias events split into 
    100 shards (named \texttt{spacepoints\_[000-099].npz}), each 
    with $\num{100000}$ events. Each file stores:
    \begin{itemize}
        \item \texttt{spacepoints}: Array of shape $(N_i, 4)$ with columns $(E, x, y, z)$, 
        where $x, y, z$ are positions (\si{\cm}) in the TPC active volume and $E$ is the 
        ionization signal (ADC units).
    \end{itemize}

    \item \texttt{labeled/}: Contains \SI{70}{\k} events with full truth information, 
    organized into three subsets:
    \begin{itemize}
        \item \texttt{train/}: 7 shards (\texttt{$*$\_000.npz} to \texttt{$*$\_006.npz}), each 
        with \SI{10}{\k} events.
        \item \texttt{validation/}: A single file per array type for \SI{13}{\k} events.
        \item \texttt{test/}: A single file per array type for \SI{7}{\k} events.
    \end{itemize}

    Each labeled subset includes four aligned arrays:
    \begin{itemize}
        \item \texttt{spacepoints}: Array $(N_i, 4)$ with $(E, x, y, z)$.
        \item \texttt{track\_ids}: Integer array $(N_i)$ that assigns each spacepoint to a 
        unique truth track identifier.
        \item \texttt{noise\_tags}: Integer array $(N_i)$ that is equal to $1$ if the matched track’s 
        transverse momentum $p_T < \mev{60}/c$ and $0$ otherwise. This flags 
        spacepoints unlikely to reach the TPC active volume. 
        \item \texttt{pid\_labels}: String array $(N_i)$ that is derived from Particle Data Group~\cite{ParticleDataGroup:2024cfk} codes and grouped 
        into five categories (labels): \emph{charged pion} ($1$), \emph{charged kaon} ($2$), 
        \emph{proton/antiproton} ($3$), \emph{electron/positron} ($4$), and \emph{others} ($0$).
    \end{itemize}

    \item \texttt{scripts/}: Contains Python utilities for data handling:
    \begin{itemize}
        \item \texttt{demo.ipynb}: Example code for data loading and visualization.
        \item \texttt{plot.py}: Helper functions for data visualization.
    \end{itemize}
\end{itemize}

Each event includes reconstructed TPC spacepoints, as well as (for the labeled set) truth associations and per-track kinematics. As shown in \autoref{fig:app-dist}, event complexity ranges from a few hundred to several thousand spacepoints and from fewer than $10$ to nearly $100$ truth tracks per event, with truth tracks spanning from fewer than $10$ to nearly $100$ per event.

{
\begin{figure}[ht]
    \centering
    \tikzsetnextfilename{fig_distribution_density}
    % \tikzexternaldisable
    \ifcompilefigures
        \resizebox{\textwidth}{!}{\input{figures/distribution_density}}
    \else
        \resizebox{\textwidth}{!}{\includegraphics{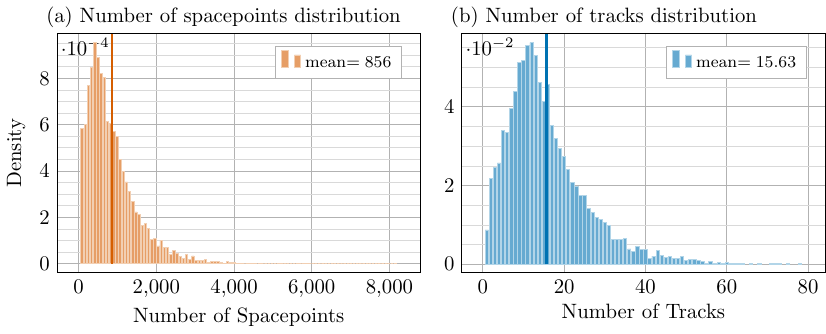}}
    \fi 
    \caption{Panel (a): Distribution of spacepoints per event. Panel (b): distribution of truth tracks per event.}
    \label{fig:app-dist}
\end{figure}
}

Given the information in the labeled dataset, several reconstruction tasks can be posed. The following defines three tasks: track finding, noise tagging, and particle identification (PID), along with the corresponding labels derivable from the provided truth.

\begin{figure}[ht]
    \centering
    \tikzsetnextfilename{fig_downstream_tasks}
    % \tikzexternaldisable
    \def\width{.5\textwidth} % increase this number to decrease the label font size
    \ifcompilefigures
        \resizebox{\textwidth}{!}{\input{figures/downstream_tasks}}
    \else 
        \resizebox{\textwidth}{!}{\includegraphics{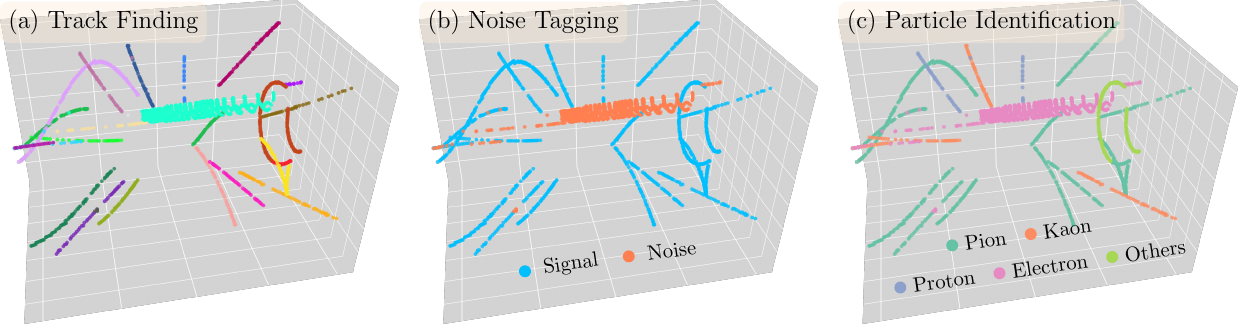}}
    \fi
    \caption{Downstream tasks ground truth labels.}
    \label{fig:downstream_tasks}
\end{figure}

\begin{figure}[ht]
    \centering
    \tikzsetnextfilename{fig_boxplot}
    % \tikzexternaldisable
    \ifcompilefigures
        \resizebox{.95\linewidth}{!}{\input{figures/boxplot}}
    \else
        \resizebox{.95\linewidth}{!}{\includegraphics{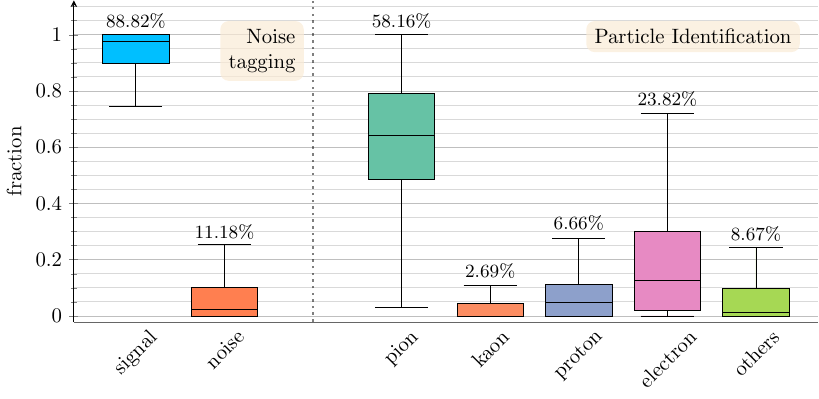}}
    \fi
    \caption{Box plots of per-event ground truth class ratios in the test subset of the labeled dataset, 
    shown for noise tagging and particle identification tasks. 
    The horizontal line within each box denotes the median per-event class ratio, the box edges indicate the interquartile range, 
    and the whiskers show the full range across events. The dataset-wide average fractions are annotated above the whiskers.}
    \label{fig:app-class_ratio}
\end{figure}

\begin{description}
  \item[Track finding:] assign every reconstructed spacepoint to a track identity using the provided truth associations. The label for each spacepoint is the integer \texttt{track\_id} of its matched truth track. \textit{Note}: Evaluation may be restricted to primaries or tracks passing kinematic cuts, but the labels themselves are unchanged.

  \item[Noise tagging:] binary classification noting whether or not a spacepoint is noise. 

  \item[PID:] multiclass classification of truth tracks into the five most commonly seen charge–conjugate categories: $\pi^\pm$ (charged pion), $K^\pm$ (charged Kaon), $p/\bar{p}$ (proton/antiproton), $e^\pm$ (electron/positron), and all other particles.
\end{description}
Figure~\ref{fig:downstream_tasks} shows an example event display with all downstream labels.
The labels are constructed directly from the provided truth associations and per-track kinematics. Class ratios for the noise and PID tasks in the labeled subset are shown in \autoref{fig:app-class_ratio}.

% (Optional) Suggested evaluation metrics for users:
\subsection*{Evaluation Metrics}
The dataset includes recommended evaluation metrics for the supported machine learning tasks. For noise tagging and PID, standard classification metrics (accuracy, precision, and recall) are recommended. 

For track finding, double-majority matching~\cite{ju2021performance} is suggested to assess efficiency and purity, following the terminologies and conventions proposed in~\cite{ju2021performance} and~\cite{garciageometric}. 

Given a predicted track candidate labeled by $t$ and a ground truth particle with label $p$, we define:
\begin{itemize}
    \setlength{\itemsep}{0.1em}
    \item $c(t, p)$ as the number of hits that have predicted label $t$ and true label $p$
    \item $c(:, p)$ as the number hits with true label $p$
    \item $c(t, :)$ as the number of hits with predicted label $t$
\end{itemize} 
and
\begin{itemize}
    \setlength{\itemsep}{0.1em}
    \item \textbf{Hit efficiency} as $c(t, p) / c(:, p)$
    \item \textbf{Hit purity} as $c(t, p) / c(t,:)$.
\end{itemize}

To match a track candidate to a particle, we follow the \textbf{double-majority rule} wherein the track candidate $t$ is matched to particle $p$ if -- and only if -- both hit efficiency and purity exceed $50\%$. This stringent rule guarantees that a track candidate can be matched to at most one particle, and a particle can also be matched by no more than one track candidate.

After the matching is done, the metrics that evaluate the overall tracking performance can be defined. Let $T$ be the number of track candidates, $P$ be the number of ground truth particles, and $M$ be the number of matches. The efficiency and purity of tracking can be defined as:
\begin{itemize}
    \setlength{\itemsep}{0.1em}
    \item \textbf{Tracking efficiency} as $M/P$ 
    \item \textbf{Tracking purity} as $M/T$.
\end{itemize}

% --------------------------
% EXPERIMENTAL DESIGN, MATERIALS AND METHODS
% --------------------------
\section*{Experimental Design, Materials, and Methods}
\noindent\textbf{Event generation and transport.}
Particles in minimum-bias \ppcol collision events at $\sqrt{s}=\gev{200}$ are generated with \pythia~8.307~\cite{m:pythia8} (Detroit tune~\cite{PYTHIA_tune}). The particles are transported with \geant~\cite{m:g4} using the FTFP\_BERT\_HP physics list via the as-built \sphenix geometry and the measured $1.4$~T solenoidal field. The simulation includes continuous energy loss, multiple scattering, secondary particle production, and decay processes when interacting with the detector material.

\vspace{2mm}
\noindent\textbf{TPC response, digitization, and reconstruction.}
Ionization signals are simulated and digitized with realistic channel-dependent gain and noise, including shaping and zero suppression. The resulting digitized 3D TPC hits are subsequently clustered into spacepoints using sPHENIX clustering algorithms. The spacepoints serve as the fundamental units for pretraining and downstream tasks, such as track finding and PID.

\vspace{2mm}
\noindent\textbf{Data preprocessing.}
All spacepoints belonging to tracks that contain fewer than five spacepoints are removed because they are too short to represent a physically meaningful particle trajectory in the TPC. Such short segments almost always originate from detector noise or random hit combinations rather than from true charged particles. 
In addition, diffractive collision processes, similar to elastic scattering, produce only a small number of particles near the beam direction and result in little activity in the central region covered by the TPC, making them uninformative for assessing central tracking performance. After removing short tracks, the number of remaining spacepoints are counted, and the event is rejected if it contains fewer than $20$ spacepoints.
%Diffractive processes, much like elastic scattering, produce only a few particles close to the beam direction and leave little activity in the central transverse region covered by the TPC, making them uninformative for evaluating central tracking performance. Therefore, after removing short tracks, we count the number of remaining spacepoints and exclude the event if it contains fewer than $20$ spacepoints.

\vspace{2mm}
\noindent\textbf{Code and reproducibility.}
All of the code used to drive the \pythia event generator and \geant detector simulations, as well as the downstream emulation and reconstruction chain (digitization; spacepoint reconstruction), is sourced from the \sphenix software stack—core simulation/reconstruction libraries and supporting infrastructure:~\citep {sphenix_coresoftware,sphenix_acts,sphenix_macros,sphenix_calibrations}. For exact reproducibility, we provide repository URLs and the commit hashes used to generate this release, as well as the corresponding \sphenix analysis build release number (ana.435) from the CernVM File System (CVMFS)~\citep{Blomer:2011tgq}, which encapsulates the full build environment.
%the same hashes are referenced in the repository \texttt{README.md} and in a machine-readable manifest (\texttt{reproducibility.yml}). Runs are deterministic given the supplied configuration files and random seeds.

% --------------------------
% LIMITATIONS
% --------------------------
\section*{Limitations}
The dataset is simulation-based and reflects the detector, material, and electronics models used. Residual differences from real data may exist.

% --------------------------
% ETHICS STATEMENT
% --------------------------
\section*{Ethics Statement}
The authors confirm this work does not involve human subjects, animal experiments, or data collected from social media platforms and follows the ethical requirements for publication found in \textit{Data in Brief}.

% --------------------------
% ACKNOWLEDGEMENTS
% --------------------------
\section*{Acknowledgments}
The authors would like to express their sincere gratitude to the sPHENIX Collaboration for sharing the simulation data and experimental knowledge, as well as Jubin Choi, Abhay Deshpande, Alexei Klimentov, Michael Begel, Torre Wenaus, Nicholas D'Imperio, James Dunlop, and John Hill from Brookhaven National Laboratory for their valuable support and feedback.

This work was supported by the Laboratory Directed Research and Development (LDRD) Program at Brookhaven National Laboratory (LDRD 25-045). Brookhaven Lab is operated and managed for the U.S. Department of Energy (DOE) Office of Science (SC) by Brookhaven Science Associates under contract No. DE-SC0012704.
Shuhang Li was partially supported by DOE-SC through the Office of Nuclear Physics under Award No.~DE‐FG02‐86ER40281.
Yihui Ren, Xihaier Luo, and Shinjae Yoo were partially supported by DOE-SC through the Office of Advanced Scientific Computing Research and the Scientific Discovery through Advanced Computing (SciDAC) program.
%under Award No.~\ray{TODO double check 9233218CNA000001}.

This research also utilized resources of the National Energy Research Scientific Computing Center (NERSC) under the ``GenAI@NERSC'' program. NERSC is a DOE Office of Science User Facility with Award No. DDR-ERCAP0034059. The authors are grateful to the NERSC staff for their support, particularly Shashank Subramanian and Wahid Bhimji.

% --------------------------
% DECLARATION OF COMPETING INTERESTS
% --------------------------
\section*{Declaration of Competing Interests}
The authors declare they have no known competing financial interests or personal relationships that could have appeared to influence the work reported in this paper.

\clearpage
% --------------------------
% REFERENCES
% --------------------------
\bibliographystyle{elsarticle-num}
\bibliography{bib} % ensure bib has entries for PYTHIA, Geant4, \sphenix software, etc.

\end{document}